\documentclass[notitlepage,aps,prx,twoside,superscriptaddress,showkeys,nofootinbib]{revtex4-1}
\usepackage{graphicx}
\usepackage{amsmath,amsfonts,amssymb,amsthm,amsbsy,mathtools}
\usepackage{bbold}
\usepackage{epic}
\usepackage{eepic}
\usepackage{mathrsfs}
\usepackage[all]{xy}
\usepackage{lmodern}
\usepackage[T1]{fontenc}
\usepackage[utf8]{inputenc}
\usepackage{color}

\newcommand{\ket}[1]{\mathop{\left|#1\right>}\nolimits}            

\newcommand{\ketbra}[2]{| #1\rangle\!\langle #2 |}
\newcommand{\dketbra}[2]{| #1\rangle \! \rangle\!\langle \! \langle #2 |}
\newcommand{\dket}[1]{|#1\rangle \! \rangle}
\newcommand{\dbra}[1]{\langle \! \langle #1 \vert}

\newcommand{\Tr}[1]{\mathop{{\mathrm{Tr}}_{#1}}}            

\def\N{\mathcal{N}}

\def\H{\mathcal{H}}
\def\M{\mathcal{M}}

\def\U{\mathcal{U}}

\def\A{\mathcal{A}}

\def\S{\mathcal{S}}
\def\V{\mathcal{V}}
\def\T{\mathcal{T}}

\newmuskip\pFqmuskip   

\newcommand*\pFq[6][8]{%
  \begingroup 
  \pFqmuskip=#1mu\relax
  \mathcode`\,=\string"8000
  \begingroup\lccode`\~=`\,
  \lowercase{\endgroup\let~}\pFqcomma
  {}_{#2}F_{#3}{\left[\genfrac..{0pt}{}{#4}{#5};#6\right]}
  \endgroup
}
\newcommand{\pFqcomma}{\mskip\pFqmuskip}

\begin{document}
	
\title{Dynamics of quantum causal structures}

\author{Esteban Castro-Ruiz}
\author{Flaminia Giacomini}
\author{\v{C}aslav Brukner}
\affiliation{Vienna Center for Quantum Science and Technology (VCQ), Faculty of Physics,
University of Vienna, Boltzmanngasse 5, A-1090 Vienna, Austria}
\affiliation{Institute for Quantum Optics and Quantum Information (IQOQI),
Austrian Academy of Sciences, Boltzmanngasse 3, A-1090 Vienna, Austria}
	
\begin{abstract}
It was recently suggested that causal structures are both dynamical, because of general relativity, and indefinite, due to quantum theory. The process matrix formalism furnishes a framework for quantum mechanics on indefinite causal structures, where the order between operations of local laboratories is not definite (e.g. one cannot say whether operation in laboratory A occurs before or after operation in laboratory B). Here we develop a framework for ``dynamics of causal structures'', i.e. for transformations of process matrices into process matrices. We show that, under \textit{continuous and reversible} transformations, the causal order between operations is always preserved. However, the causal order between a subset of operations can be changed under continuous yet nonreversible transformations. An explicit example is that of the quantum switch, where a party in the past affects the causal order of operations of future parties, leading to a transition from a channel from A to B, via superposition of causal orders, to a channel from B to A.  We generalise our framework to construct a hierarchy of quantum maps based on transformations of process matrices and transformations thereof.
\end{abstract}
	
\maketitle
\section{Introduction}
We are used to the fact that events occur in a fixed temporal order. Given two events $A$ and $B$, either $A$ is in the causal past of $B$, $A$ is in the causal future of $B$, or they are causally disconnected (spacelike separated). Although this picture seems natural, the idea that a fixed causal structure is a fundamental ingredient of the physical world has been recently challenged. Indeed, the interplay between quantum mechanics and general relativity suggests that causality might be indefinite. Because the causal structure in general relativity is determined by a dynamical field -- the space-time metric -- and dynamical quantities can be indefinite in quantum mechanics (i.e. put in superpositions of well-defined classical values), one might expect indefiniteness with respect to the question of whether an interval between two events is timelike, null or spacelike, or even whether event $A$ is before or after event $B$. In order to describe causal structures that are both dynamical, because of general relativity, and indefinite, due to quantum theory, several authors have proposed extensions to quantum theory that do not assume a definite causal structure \cite{hardy, oeckl, ocb}.
 
\begin{figure}[h]
\centering
\includegraphics[scale=0.25]{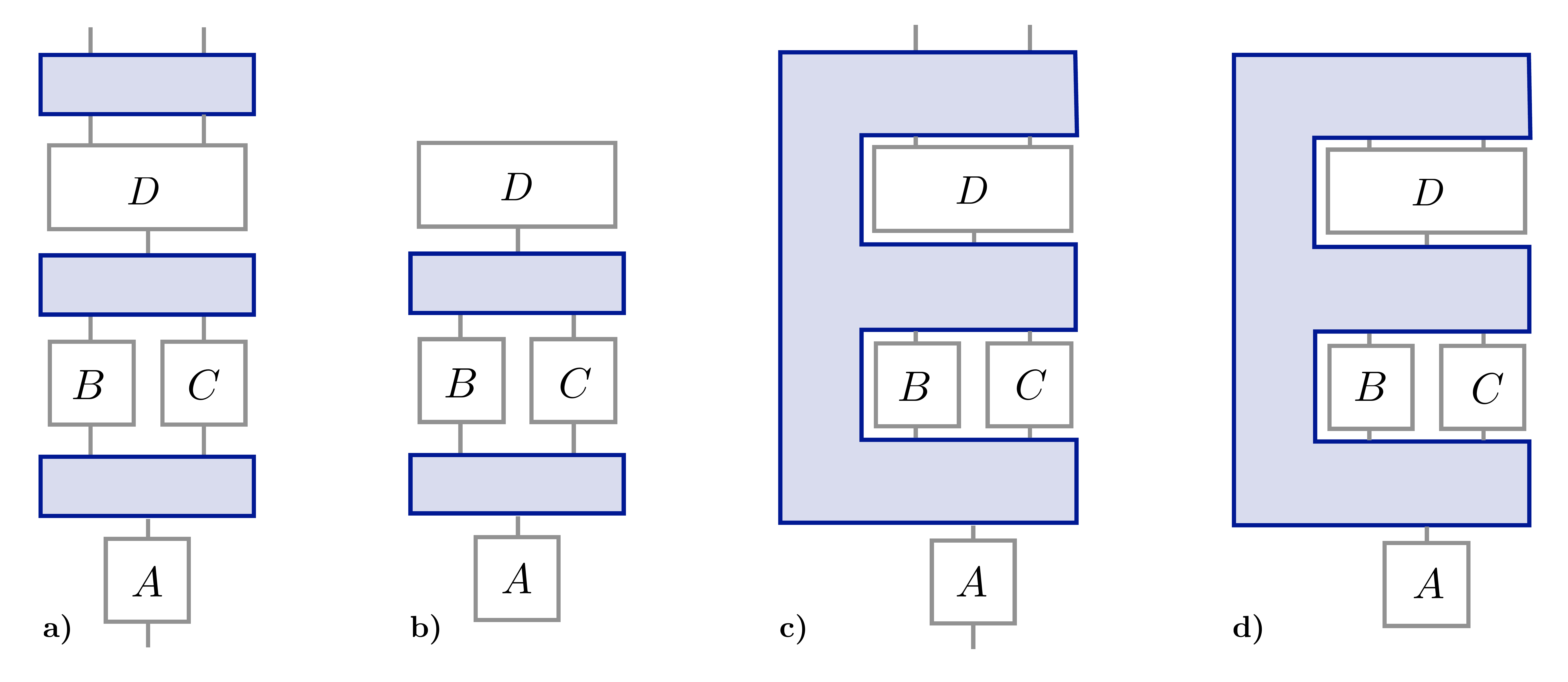}
\caption{\label{noopenends}Composition of local operations. $A$, $B$, $C$ and $D$ are local laboratories where the usual quantum formalism holds. The wires entering and leaving the laboratories represent quantum systems on which operations are applied. \textbf{a)} Local operations are composed to form a system with definite causal structure and open ends (wires whose ends are not attached to any box). Information flows from bottom to top and the wiring determines the order of operations but is not enough to determine the probabilities for outcomes of local experiments, because these may change under the influence of external systems through the past open end. \textbf{b)} A quantum circuit: Local operations are composed with a causally ordered process to yield an object with no open ends. The probabilities for local experiments are determined completely by the choice of local operation and the wiring between boxes. \textbf{c)} Composition of local operations with indefinite causal structure. The order of application of operations $B$ and $C$ is not defined \textit{a priori}. Due to the open end in the past, the causal order and the probabilities depend in general on an external system entering through the open end. \textbf{d)} The composition of local operations with a process matrix yields an object with indefinite causality and no open ends. The probabilities are determined completely by the choice of local operation and the process matrix (depicted by the blue E-shaped object) connecting the boxes. The object obtained from the composition is the generalisation of a quantum circuit for the case where the causal order is indefinite.}
\end{figure}

The process matrix formalism \cite{ocb, witness, oreshkov, infinitedimensions} achieves such a goal. Its central notion is that of a ``process'' (or a ``process matrix''), which is a generalisation of the notion of a ``physical state'' (degrees of freedom over a spacelike hypersurface) and of a ``channel'' (degrees of freedom over a timelike hypersurface). The operational framework in which process matrices are defined is depicted in Fig. (\ref{noopenends}). In the framework, observers perform experiments in their local laboratories, where quantum mechanics is assumed to hold. The process matrix is the object that ``wires'' or ``connects'' local operations together, specifying the causal order (definite or indefinite) between such operations. Roughly speaking, it accounts for all the physical processes outside the local laboratories. Process matrices composed with the local operations form ``closed'' systems. This means that the probabilities for the events described in the laboratories are \textit{completely} determined by the choice of local operations performed by the parties and the way the process matrix connects the laboratories. In a closed system, the probabilities do not depend on any ``external'' operation. Because of this independence, this definition of a closed system matches the usual notion of a closed system in physics, in the sense that there is no physical interaction between the system and everything external to it. We say that a composition of operations has no ``open ends'' if it is a closed system. A process matrix connected with local operations generalises the notion of a quantum circuit \cite{hardyoperatortensor}, and reduces to it in the case where the causal order between the laboratories is fixed.

A process is called \textit{causally separable} if it can be written as a probabilistic (convex) mixture of states and channels. In the most general scenario, whether a party $A$ can signal to a party $B$ might depend on the choice of operation of yet another party $C$, and a general notion of causal nonseparability is required to incorporate these cases \cite{oreshkov, abbot}. Causally non-separable processes can give rise to correlations that can violate ``causal inequalities'', which are satisfied if the events are ordered according to a fixed causal order \cite{oreshkov, abbot}. This is a direct analogy to the violation of Bell’s inequalities by quantum correlations, which are satisfied if the correlations fulfil the condition of local causality \cite{bell}. While we still lack an understanding as to whether there are correlations in nature that can violate causal inequalities, we do know that physically implementable causally non-separable processes exist. One particular case is the ``quantum switch'', an auxiliary quantum system that can coherently control the order in which operations are applied. The quantum-switch technique offers the possibility to implement certain information-theoretical tasks that quantum circuits with a fixed order of operations cannot perform \cite{chiribellaswitch, araujo, allard}. For this reason, it is important to understand how these processes, as useful quantum information resources, can be obtained \textit{dynamically}, e.g. from separable process matrices. 

We have achieved significant progress in the characterisation of quantum causal structures \cite{ocb, witness, oreshkov}, yet we still lack a theory of their dynamics, intended as transformations that change the way information is transmitted. What is the most general transformation that takes a process matrix as an input and returns another process matrix as an output? Are there transformations that take a process matrix we are able to interpret physically (say, a quantum channel from A to B) and outputs a process matrix that violates a causal inequality? A negative answer would provide a strong argument in favour of the view that no physical processes can violate causal inequalities. A positive one would be even more remarkable, as it would indicate a way how to physically realise a process that could violate the strongest notion of causality.

In this paper we develop a theory of ``dynamics of process matrices'', giving a full characterisation of ``supermaps'', that map process matrices into process matrices. We extend the theory of transformations further by including transformations not only from process matrices to process matrices, but also of the transformations thereof, constructing an infinite hierarchy of transformations. Our approach is similar to that of Ref. \cite{perinotti}, where a general classification of higher order quantum computations is given.

We prove that, under continuous (i.e. continuously connected to the identity) and reversible transformations of process matrices, the causal structure \textit{cannot be changed}, because all such transformations are local unitary operations in the parties’ input and output Hilbert spaces. In other words, under continuous and reversible dynamics the causal order between local operations is preserved. However, there exist processes in which a party $A$ can apply continuously parametrised reversible transformations in his/her local laboratory in such a way that the \textit{reduced} process, that is, the process for the remaining parties obtained after applying $A$'s operation to the original process, exhibits different causal structures for different values of the parameter. We construct an explicit example of this fact for the case of the quantum switch, in which a party in the past controls the causal order of parties in the future.

A consequence of our results is that, starting from a process with definite causal order, one cannot arrive at one that displays indefinite causal order in a continuous and reversible fashion. If one takes the view that physical transformations are continuous and reversible, our results mean that no process that violates causal inequalities can be obtained from physically realisable causal structures with a definite causal order -- states and channels. Assuming that transformations are reversible but might not even be continuous, we prove a result in favour of the conjecture that the original process from Ref. \cite{ocb} is not physically realisable, by showing that it cannot be ``reached'' from any causally separable process via a reversible map. 

\section{Process matrices}
\label{processmatrices}

The operational idea of a causal influence is best illustrated by considering two observers, $A$ and $B$, performing experiments in two separate laboratories. At each run of the experiment, the observers receive a physical system only once and perform an operation on it. Afterwards, the observers send the system out of the laboratory.  Each laboratory features a device with an input and an output (see Figure \ref{noopenends} for the multipartite case), that outputs an outcome (the result of the experiment), for a given choice of the input (the knob settings determining which experiment is performed). If the experiments performed in the local laboratories are described by quantum theory, to each observer corresponds a Hilbert space $\H^X$, for $X = A, B$. It is convenient to split $\H^X$ into input and output subspaces, that is $\H^X = \H^{X_I} \otimes \H^{X_O}$, where $I$ stands for input and $O$ for output and $X = A, B$ labels the local observer. Local operations are represented by quantum instruments, that is, collections of completely positive (CP) maps $\{\M^A_{i}\}$ for $A$ and $\{\M^B_{j}\}$ for $B$ that map systems from the input Hilbert space of each party to the corresponding output Hilbert space, yielding outcomes labeled by $i$ and $j$. The conservation of probability in, say, $A$'s local laboratory means that the sum over $i$ of her CP maps adds up to a completely positive, trace preserving (CPTP) map. The probability for a pair of outcomes $i$ and $j$ for a choice of knob settings given by the instruments $\{\M^A_{i}\}$ and $\{\M^B_{j}\}$ is a bilinear function of the local CP maps. Using the Choi-Jamio\l kowski (CJ) representations \cite{jamiolkowski, choi} of the local CP maps, whereby the CP map $\N: \mathcal{L}(\H^{A_I}) \longrightarrow \mathcal{L}(\H^{A_O})$ corresponds to the positive-semidefinite operator $C_\N \in \mathcal{L}(\H^{A_I}\otimes\H^{A_O})$ given by $C_\N = ( \mathbb{1}^{A_I} \otimes \N)\dket{\mathbb{1}}\dbra{\mathbb{1}}$, where $\dket{\mathbb{1}} = \sum_i \ket{ii} \in \H^{A_I}\otimes\H^{A_I}$, the joint probability for the outcomes $i$ and $j$ can be expressed as 
\begin{equation}
\label{brule}
p_{ij} = \Tr{}\left(W C_{\M^A_{i}} \otimes C_{\M^B_{j}}\right),
\end{equation}
where W is a ``process matrix'' that describes the causal structure outside of the laboratories. Mathematically, it is a positive semidefinite operator that connects the two laboratories by acting on the tensor product of the input and output Hilbert spaces of $A$ and $B$. The set of valid process matrices is defined by the requirement that probabilities are well defined – that is, they must be non negative and add up to 1. These requirements are equivalent to the following constrains:
\begin{subequations}
\label{wconditions}
\begin{align}
W & \geq 0 \label{wcondition1}\\
\Tr{}\left(W\right) &= d_{O} \label{wcondition2}\\
P(W) &= W \label{wcondition3},
\end{align}
\end{subequations}
where $d_{O}$ is the dimension of the output Hilbert space and $P$ is a self-adjoint real projection operator. This operator, first specified in \cite{witness}, is written explicitly in Appendix A.
Physically, conditions (\ref{wconditions}) exclude Deutsch’s \cite{deutsch} and Lloyd’s \cite{lloyd} closed time-like curves or any causal loops that would allow a party to send a signal into her/his past and hence give rise to the so-called ``grandfather paradox'' \cite{oreshkov, allen, baumeler}. 

Process matrices, i.e. matrices $W\in \mathcal{L(H)}$ satisfying (\ref{wconditions}) contain quantum states and quantum channels as particular cases. Concretely, a process matrix of the form $W = \rho^{A_I B_I} \otimes \mathbb{1}^{A_O B_O}$, where $\rho^{A_I B_I}$ is a unit trace, positive matrix, represents the situation in which $A$ and $B$ share a quantum state. A process matrix $W = W^{A_I B_I A_O} \otimes \mathbb{1}^{B_O}$, where $W^{A_I B_I A_O}$ is a matrix such that conditions (\ref{wconditions}) are satisfied, represents a channel (possibly with memory) from $A$ to $B$. A channel from $B$ to $A$ has an analogous form but with $A$ and $B$ interchanged. A bipartite process matrix $W$ is called \textit{causally ordered} if it is either a state shared by $A$ and $B$, a channel from $A$ to $B$ or a channel from $B$ to $A$. We can also think of situations in which the process has a definite causal order, but for which only probabilistic predictions can be made regarding which causal order is realised. To capture this situation, we define a process to be \textit{causally separable} if it is a probabilistic (convex) mixture of causally ordered processes. 

An example of a causally non-separable process is the quantum switch. In the switch, two parties, $A$ and $B$, act on a target quantum system in an order which is coherently controlled by another quantum system. In order to account for the target and control quantum systems as well as for the parties $A$ and $B$, the quantum switch is formally a tripartite process matrix, with the third party, $C$, being always in the future of $A$ and $B$. It is reasonable to extend the definition of causal separability for this case in exactly the same way as for the bipartite case, precisely because $A$ and $B$ can always signal to $C$ but $C$ cannot signal to them. We say that a process is causally separable if it can be written as a probabilistic mixture of processes in which ``$A$ signals to $B$ and $B$ signals to $C$'' or ``$B$ signals to $A$ and $A$ signals to $C$''. (Note that when we write, say, ``$A$ signals to $B$ and $B$ signals to $C$''  it is implicitly assumed that signalling from $A$ to $C$ is also possible.) 

The complete Hilbert space for the quantum switch is therefore $\mathcal{H} = \mathcal{H}^{A_I}\otimes\mathcal{H}^{B_I}\otimes\mathcal{H}^{A_O}\otimes\mathcal{H}^{B_O}\otimes\mathcal{H}^{C_I}$. The input Hilbert space of $C$ is divided in target and control spaces, $\mathcal{H}^{C_I} =  \mathcal{H}^{C_T}\otimes\mathcal{H}^{C_C}$. The output Hilbert space of $C$ is trivial, $\mathcal{H}^{C_O} = \mathbb{C}$, because $C$  receives and measures a system but does not prepare one. The quantum switch is then written as $W_{S} = \ketbra{S}{S}$, where
\begin{equation}
\label{switch}
\ket{S} = \frac{1}{\sqrt{2}}\ket{ABC}^{ABC_T}\ket{0}^{C_C} + \frac{1}{\sqrt{2}}\ket{BAC}^{ABC_T}\ket{1}^{C_C},    
\end{equation} 
with $\ket{XYC}^{ABC_T} = \ket{\psi}^{X_I}\dket{\mathbb{1}}^{X_O Y_I}\dket{\mathbb{1}}^{Y_O C_T}$, for $X,Y = A, B$. 
Because it is a rank-one projector, it cannot be written as a nontrivial probabilistic mixture of other processes. The fact that it leads to signalling both from $A$ to $B$ to $C$ and from $B$ to $A$ to $C$ shows then that it is a causally non-separable process. In Section \ref{supermaps} we show how the quantum switch can be obtained from a causally ordered process via a reversible and discontinuous process matrix transformation.  

Although the quantum switch is a causally non-separable process, it was shown \cite{witness, oreshkov} that there does not exist \textit{any} causal inequality that can be violated by the quantum switch. Analogous to Bell inequalities, which bound the set of possible correlations realisable within a local hidden variable model, causal inequalities bound, in a device-independent way, the set of possible correlations realisable within models compatible with a global, fixed causal structure. Therefore, although the switch is a causally non-separable process, it does not violate the strongest, device-independent notion of causality. In contrast, the bipartite process $W_{OCB} \in \mathcal{L}( \H^{A_I} \otimes \H^{B_I} \otimes \H^{A_O} \otimes \H^{B_O})$ from \cite{ocb} defined by
\begin{equation}
\label{wocb}
W_{OCB} = \frac{1}{4}\left( \mathbb{1} + \frac{1}{\sqrt{2}}(\mathbb{1}\otimes\sigma_z\otimes\sigma_z\otimes\mathbb{1} + \sigma_z\otimes\sigma_x\otimes\mathbb{1}\otimes\sigma_z )\right),	
\end{equation}  
where $\sigma_i$ denotes the $i$th Pauli matrix, is a valid bipartite process matrix that violates a causal inequality and is thus incompatible with a pre-established global causal order. As we will see in \ref{supermaps}, there is no reversible process matrix transformation that takes a causally separable process to $W_{OCB}$.

\section{Transformations of process matrices}
\label{supermaps}
We now introduce transformations that map process matrices into process matrices. These transformations define the most general dynamics of causal structures that is compatible with the local validity of quantum theory. By ``dynamics'' we mean transformations that take a process matrix as an input and give a process matrix as an output. By definition, these transformations preserve the validity of process matrices, that is, conditions (\ref{wconditions}), regardless the input process matrix, analogous to the way quantum channels take valid quantum states to valid quantum states independently of the input state. A process matrix transformation may take a process in which $A$ signals to $B$ by means of a quantum channel, to a different process, for example, the quantum switch, where the direction of signalling between $A$ and $B$ is controlled coherently. 

Quantum states and quantum channels are special cases of process matrices, both having dynamical evolution. For the case of a quantum state, dynamics is governed by the Schr\"odinger equation, whereas quantum channels can evolve in time when the physical constituents employed for their implementation change. 

For example, the situation of a quantum system traversing an optical fibre whose properties change in time can be modelled by a time-evolving channel. The type of transformations we consider here can be seen as a generalisation of these two cases to the situation where the causal structure is indefinite. Note, however, that process  matrix transformations are maps of a completely general character and may, but need not be understood as the evolution of a process \textit{in time} (as defined by some observer). 

Our formulation of dynamics of process matrices allows to explore the question about the realisability of processes with indefinite causal order. Given a process matrix with a certain causal structure, is it possible to transform it in such a way that the output process matrix has a different causal structure? Can we transform causally separable processes into causally non-separable ones? In order to make these questions more precise, it is important to point out that process matrices, as stated in the Introduction, yield objects with no ``open ends'' when connected with all the relevant local operations, i.e. the local operations corresponding to all the parties that play a role in the process. In this way a, say, bipartite process for parties $A$ and $B$, yields an object with no ``open ends'' when connected with a local operation for $A$ and a local operation for $B$. The statement that a process yields objects with no open ends means that the probability distribution corresponding to an experiment is determined \textit{completely} by the choice of local operations made by the parties and the way these local operations are ``wired'' together. Just to be clear: The way we have defined objects, it is not the process matrix that has no open ends but rather the composition of the process matrix with the relevant local operations. For the case of quantum circuits \cite{hardyoperatortensor}, the wiring of the boxes, together with the convention that information flows from bottom to top, determines the order in which the operations are applied, see Fig. (\ref{noopenends}a) and (\ref{noopenends}b). For the more general case of an indefinite causal structure, in which the order of local operations is not fixed in principle, the wiring of the boxes is replaced by ``E-shaped'' objects (see Fig. (\ref{noopenends}c) and (\ref{noopenends}d)), that specify the causal structure (definite or indefinite) outside the local laboratories. For an object with indefinite causality and open ends, (Fig. (\ref{noopenends})c), the notion of causal non separability is in general not well defined, because the signalling between the parties is, in the general case, determined not only by their choice of local operations and the ``E-shaped'' object outside the laboratories, but also by the external systems influencing the experiment via the open ends. In some cases, the natural way to proceed in this case is to ``close'' the wires by inserting local laboratories at each open end, and study the causal (non)separability of the process matrix thus obtained. Note that this process matrix will have more parties than the original ``E-shaped'' object, and the question whether it is causally separable or not has to be analysed considering the newly added parties. 

\begin{figure}[h]

\centering
\includegraphics[scale=0.35]{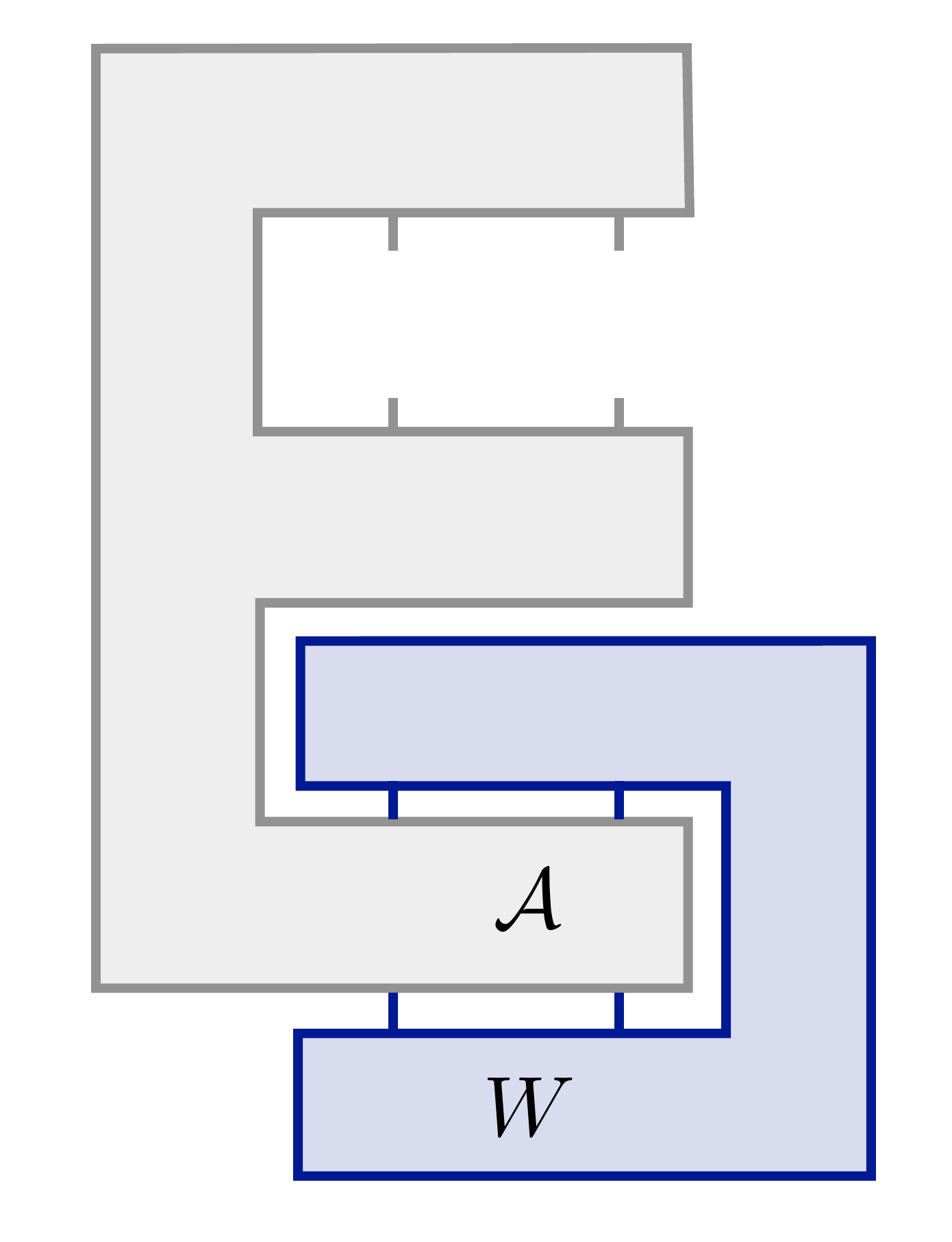}
\caption{\label{supermap}Pictorial representation of a process matrix transformation. The action of the transformation $\A$ on the bipartite process $W$ can be visualised as the composition of the objects in the picture. The ``supermap'' $\A$ transforms the process matrix $W$ by acting on its ``slots'', where local operations are applied in order to yield probabilities. The object resulting from the action of $\A$ on $W$, $\A(W)$ is a valid bipartite process matrix for all valid bipartite input process matrices $W$. The ``slots'' of the new process matrix $\A(W)$ (top part of the picture) act on local operations to give probabilities.}
\end{figure}

Given the definition of a process matrix as an operation that yields an object with no open ends, we can visualise the action of a process matrix transformation $\A$ on a process matrix $W$ in the way depicted in Fig. (\ref{supermap}). This pictorial representation, resembling the one of quantum combs \cite{chiribellacombs}, allows us to think of the \textit{evolution} of a process in terms of the composition of an input process matrix $W$ with a process matrix transformation $\A$. As we will see below, processes can in general evolve from a definite to an indefinite causal order. However, we show that this is never the case under physically justified continuous and reversible transformations. In this sense, the causal structures are ``robust'': a ``world'' that has a definite causal order will never evolve in in a world with an indefinite causal order and vice versa.

In the next subsection we characterise transformations from process matrices to process matrices.
  
\subsection{Definition of process matrix transformations}
Let $\mathcal{H}_1$ and $\mathcal{H}_2$ be Hilbert spaces and $\mathcal{L}(\mathcal{H}_1)$ and $\mathcal{L}(\mathcal{H}_2)$ be the corresponding Hilbert spaces of linear operators on $\mathcal{H}_1$ and $\mathcal{H}_2$. Let us call $\mathcal{W}_1$ and $\mathcal{W}_2$ the set of all valid process matrices in $\mathcal{L}(\mathcal{H}_1)$ and $\mathcal{L}(\mathcal{H}_2)$, respectively. Note that it is implicitly assumed that $\mathcal{H}_1$ and $\mathcal{H}_2$ have an input-output and multipartite tensor product structure as the one in Section (\ref{processmatrices}) to allow for the construction of process matrices on them. Process matrices in these spaces are characterised by the matrices $W$ that satisfy conditions (\ref{wconditions}) in their respective spaces of operators. 
	
By the definition of a process matrix, a transformation $\mathcal{A}:\mathcal{L}(\mathcal{H}_1)\longrightarrow\mathcal{L}(\mathcal{H}_2)$ is a valid process matrix transformation from the input space $\mathcal{L}(\mathcal{H}_1)$ to the output space $\mathcal{L}(\mathcal{H}_2)$ if it preserves conditions (\ref{wconditions}) for all input process matrices. Let us analyse what each of these conditions imply for $\mathcal{A}$. Condition (\ref{wcondition1}) implies that $\mathcal{A}$ is a positive map. As for the case of quantum states, it is physically meaningful to consider process matrix transformations that act on a proper subset of the parties of a multipartite process. Moreover, it is also physically sound that the complete set of parties of the transformed process can share entangled quantum systems. As we show in Appendix B, these requirements imply that the map $\mathcal{A}$ is completely positive. An alternative proof of complete positivity can be found in Ref. \cite{perinotti}.
Condition (\ref{wcondition2}) implies that $\mathcal{A}$ must be \textit{trace rescaling}, that is, it must rescale the trace of the input process matrix $d_1^{O}$ to the dimension of the space $\mathcal{H}_2^{O}$, $d_2^{O}$. Because this condition has to be satisfied for all process matrices, $\A$ must be of the form $\A = \frac{d_2^{O}}{d_1^{O}}\tilde{\A}$, where $\tilde{\A}$ is a trace preserving map. In particular, if the input space is isomorphic to the output space, then $\A$ is a completely positive trace preserving map. Requiring that $\mathcal{A}$ preserves (\ref{wcondition3}) is equivalent to requiring 
\begin{equation}
P_2\circ \mathcal{A} \circ P_1 = \mathcal{A} \circ P_1,
\label{c} 
\end{equation}
where the subindices in the projector $P$ of (\ref{wcondition3}) indicate that the input and output space may be different. This general requirement captures not only the case where the dimension of the Hilbert spaces in which the parties act changes after the transformation, but also the case where the number of parties is different before and after the transformation. 
 
In order to relate the conditions implied for $\mathcal{A}$ and conditions (\ref{wconditions}), it is convenient to rewrite the conditions for $\mathcal{A}$ in terms of its CJ operator, $C_{\mathcal{A}}$. The first two conditions for $\mathcal{A}$ can be rewritten in a straightforward way using the well-known properties of the CJ operator (see, for example, \cite{heinosaari}): 
\begin{itemize}
\item 
$\mathcal{A}$ is completely positive if and only if 
$C_{\mathcal{A}} \geq 0$
\item
$\mathcal{A}$ is trace-rescaling if and only if 
$\Tr{2}\left(C_{\mathcal{A}}\right) = \frac{d_2^{O}}{d_1^{O}} \mathbb{1}_1$,
\end{itemize} 
where the subindices 1 and 2 refer, respectively, to the Hilbert spaces $\mathcal{H}_1$ and $\mathcal{H}_2$. 

In order to write the third condition in terms of $C_{\mathcal{A}}$, we take the CJ operator of the right hand side of (\ref{c}). Using the fact that $\mathbb{1}\otimes \mathcal{N} (\dket{\mathbb{1}}\dbra{\mathbb{1}})= \mathcal{N}^T\otimes \mathbb{1} (\dket{\mathbb{1}}\dbra{\mathbb{1}})$ (the superscript $T$ denotes transpose with respect to the basis in which $\dket{\mathbb{1}} \in \mathcal{H}_1\otimes\mathcal{H}_1$ is defined), we find
	\begin{align}
	\mathbb{1} \otimes \mathcal{A}\circ P_1 (\dket{\mathbb{1}}\dbra{\mathbb{1}}) 
	=& P_1^T \otimes \mathbb{1} (C_{\mathcal{A}}).
	\end{align}

	On the other hand, taking the CJ of the left hand side of (\ref{c}) gives
	\begin{equation*}
	 \mathbb{1} \otimes P_2\circ \mathcal{A}\circ P_1 (\dket{\mathbb{1}}\dbra{\mathbb{1}}) = P_1^T \otimes P_2(C_{\mathcal{A}}).
	\end{equation*}	
 As noted in Section \ref{processmatrices}, and as can be explicitly seen in Appendix A, $P_1$ is a real projector. Being hermitian, it equals its transpose, $P_1 = P_1^T$. Therefore we can rewrite condition (\ref{c}) as
	\begin{equation}
	\label{ccc}
	P_1 \otimes \mathbb{1} (C_{\mathcal{A}}) =   P_1\otimes P_2 (C_{\mathcal{A}})	
	\end{equation}	
We have then written all conditions for $\mathcal{A}$ in terms of its CJ, $C_{\mathcal{A}}$. In summary, we have characterised all the possible transformations that take \textit{any} valid process matrix to another valid process matrix in terms of their CJ representation. Process matrix transformations $\A$ have CJ representations that satisfy the following:
\begin{subequations}
	\label{Aconditions}
	\begin{align}
		C_\A \geq & 0 \\
		\mathrm{Tr}_2 \left(C_\A\right) = &\frac{d_2^{O}}{d_1^{O}} \mathbb{1}_1 \\
		P_1 \otimes \mathbb{1} (C_{\mathcal{A}}) = &  P_1\otimes P_2 (C_{\mathcal{A}}).
	\end{align}
\end{subequations}

\subsection{Higher order maps}
In this section we show how the idea of process matrix transformations can be generalised to the case of higher order transformations, thereby extending the framework of process matrices, which, as we show in this Section, can be seen as transformations of order 1. First, we note that condition (\ref{ccc}) can be written as 
\begin{equation}
\label{ordertwoprojector}
(\mathbb{1} \otimes \mathbb{1} - P_1 \otimes \mathbb{1} + P_1\otimes P_2) ( C_\A) = C_\A.
\end{equation} 	
This is convenient because condition (\ref{ccc}) is rephrased in terms of a projection operator that leaves $C_\A$ invariant. More precisely, the supermap $\mathcal{A}$ has an input space $\mathcal{H}_1$ and an output space $\mathcal{H}_2$. Its CJ operator $C_\A \in \mathcal{L}(\mathcal{H}_1\otimes\mathcal{H}_2)$ is invariant under the projector $P^{(2)}_{12}:= \mathbb{1} \otimes \mathbb{1} -P_1 \otimes \mathbb{1} + P_1\otimes P_2$. Condition (\ref{ordertwoprojector}) is a generalisation of condition (\ref{wcondition3}) for process matrices. More concretely, we note that any process matrix $W$ can be considered to be the CJ of a supermap with a trivial input space $\mathcal{H}_1 = \mathbb{C}$. To see that this is the case we rewrite the condition (\ref{ccc}) for $W$. We get
	\begin{equation}
	(1 \otimes \mathbb{1} -1 \otimes \mathbb{1} + 1\otimes P) (W) =  P (W) = W,
	\end{equation}  
	which is just the original condition (\ref{wcondition3}) we know for $W$. In this sense, $W$ is a particular type of supermap, with trivial input space.
	
We can now generalise the projector (\ref{ordertwoprojector}) to build a hierarchy of higher-order transformations, in the spirit of Ref. \cite{perinotti}. We define $P^{(1)}_1 = P_1$, $P^{(1)}_2 = P_2$ and 
	\begin{equation}
	P^{(n)}_{12} = \mathbb{1}\otimes\mathbb{1} - P^{(n-1)}_1 \otimes  \mathbb{1}+ P^{(n-1)}_1\otimes P^{(n-1)}_2.	
	\end{equation} 	
From this point of view a process matrix $W =: W_1$ is a supermap of order 1 in the sense that it satisfies $P_1(W_1) = W_1$ (together with the positivity and trace-preserving condition). A map $\A$ that takes a process matrix to a process matrix is a supermap of order 2 in the sense that its CJ, $C_\A$ satisfies $P_2(C_\A) = C_\A$ (together with the positivity and trace-rescaling condition). In this way we can define these type of maps for arbitrary order, that is, a supermap of order $n$, $W_n \in \mathcal{W}_n$ is an operator that satisfies $P^{(n)}_{12}(W_n) = W_n$ (together with the positivity and trace-rescaling condition). It takes a valid supermap of order $n-1$, with corresponding Hilbert space $\H_1$, to a valid supermap of order $n-1$ with corresponding Hilbert space $\H_2$.
 
\section{Continuous and reversible transformations preserve the causal order of a process}
\label{continuousandreversible}
In this Section we formulate and state the main result of our paper, namely, that the causal structure of a process remains invariant under continuous and reversible dynamics. Consider a bipartite process matrix with a corresponding Hilbert space $\H = \H^{A_I} \otimes\H^{A_O} \otimes \H^{B_I}\otimes\H^{B_O}$. Let $\U: \mathcal{L}(\H)\longrightarrow\mathcal{L}(\H)$ be a continuous, i.e. continuously connected to the identity, and reversible transformation from process matrices to process matrices. Concretely, the continuity and reversibility requirements for $\U$ mean that there exists a one-parameter family of valid, reversible process matrix transformations  $\U_\lambda$, such that $\U_{\lambda = 0}$ is the identity transformation and $\U_{\lambda = 1} = \U$. Since, by assumption, $\U_\lambda$ is a valid process matrix transformation for all $\lambda$, it is a CPTP map for all $\lambda$. Moreover, since $\U_\lambda$ has an inverse for all $\lambda$ and the inverse is also a valid process matrix transformation, there exists a family of unitary matrices $U_\lambda = \mathrm{e}^{-\mathrm{i}\lambda H}$, for some traceless, hermitian operator $H$, such that $\U_\lambda(X) = \U_\lambda X \U_\lambda^\dagger$ for all $X\in \mathcal{L}(\H)$. As a consequence, the condition for $\U(\lambda)$ to be a valid process matrix transformation is $P(\mathrm{e}^{-\mathrm{i}\lambda H}W\mathrm{e}^{\mathrm{i}\lambda H}) =\mathrm{e}^{-\mathrm{i}\lambda H}W\mathrm{e}^{\mathrm{i}\lambda H}$ for all $W\in \mathcal{W}$, where we have omitted the subindices in the operator $P$ because the input and output Hilbert spaces are isomorphic. If we take $\lambda$ to be infinitesimal, the condition above reads $P\left(W-\mathrm{i}\lambda\left[H,W\right]\right) = P\left(W\right)-\mathrm{i}\lambda P\left(\left[H,W\right]\right) = W -\mathrm{i}\lambda P\left(\left[H,W\right]\right) = W -\mathrm{i}\lambda \left[H,W\right]$, where we have used the fact that $W$ is a valid process matrix. From this we conclude that 
\begin{equation}
\label{condition}
P(\left[H,W\right]) = \left[H,W\right]. 
\end{equation}

\begin{figure}[h]
\centering
\includegraphics[scale=0.35]{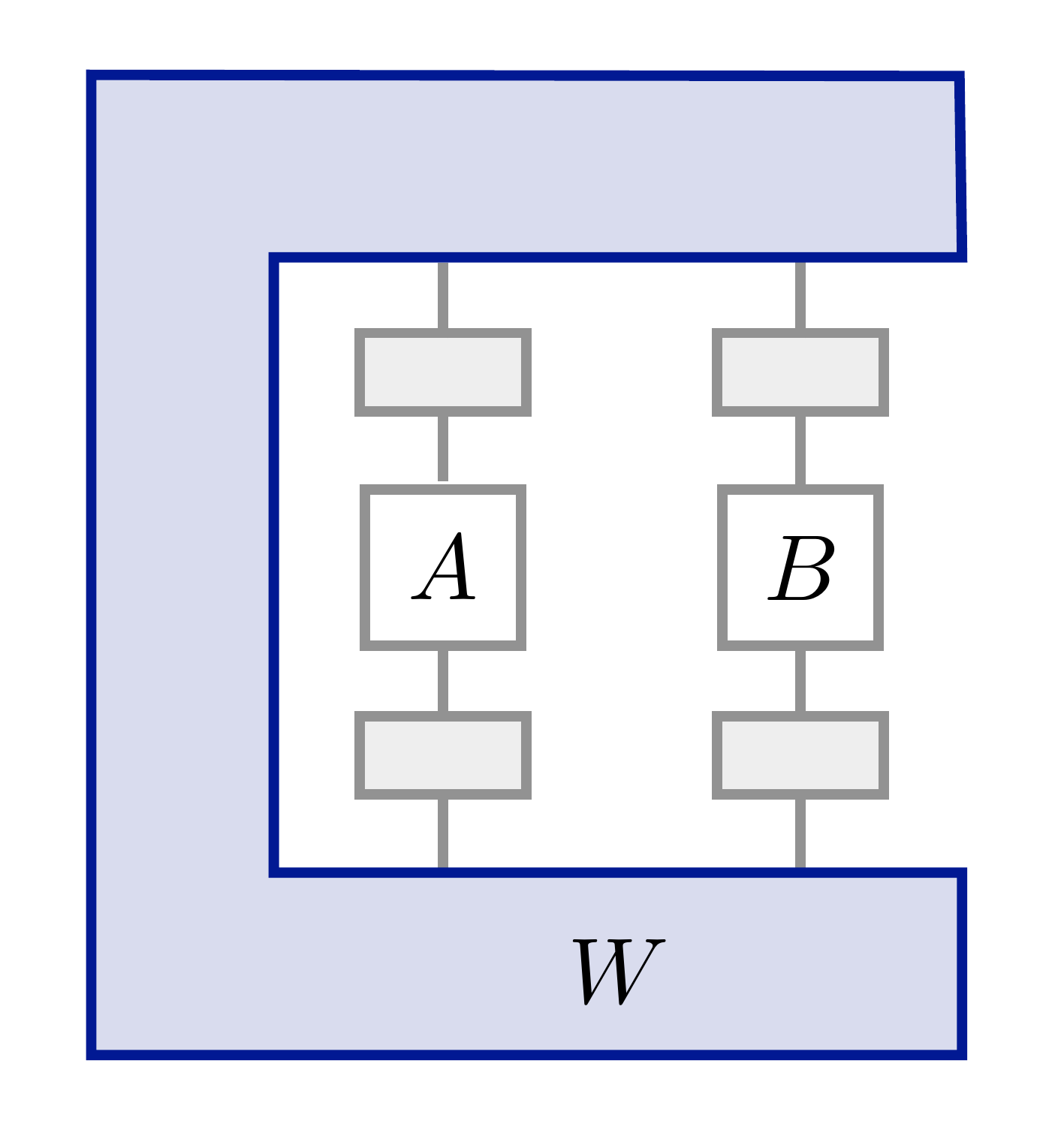}
\caption{\label{robustcausalorder}All continuous and reversible process matrix transformations are of the form depicted, for the bipartite case, in this picture. Because all of these transformations amount to local unitary operations in each party's input and output Hilbert space, continuous and reversible transformations always preserve the causal order.}
\end{figure}


In Appendix \ref{appendix1}, we show that condition (\ref{condition}) can only be satisfied by local unitaries, that is, unitaries of the form 
\begin{equation}
\label{localunitaries}
U = U_A^{I}\otimes U_B^{I}\otimes U_A^{O}\otimes U_B^{O},
\end{equation}
and we generalise the proof for an arbitrary number of parties and dimensions of the local laboratories' systems.

In conclusion, if $W$ is any process matrix and $\U$ is a continuous and reversible process matrix transformation, in the sense specified above, then $\U$ is merely a product of local unitaries, of the form of Eq. (\ref{localunitaries}), see Fig. (\ref{robustcausalorder}). Importantly, this fact implies that the causal structure of $W$ remains unchanged under the action of $\U$. This result points to the idea that causality is a robust feature of physical processes: any transformation that changes the causal order of a process must be either discontinuous or irreversible. 

\section{Examples}
\label{examples}
Let us now consider some interesting examples of process matrix transformations. 
\subsection{Maps that trivially change the causal structure}
The formalism for transformations of process matrices developed in the previous section allows trivially for transformations that map a process matrix with a given causal order (for example one in which $A$ can signal to $B$) to another process matrix in which the causal order might be different (for example one in which $B$ signals to $A$) or even indefinite, like the one in the process given in Eq. (\ref{wocb}). In fact, given any process matrix $\tilde{W}$, we can define a constant map $\A$ such that $\A(W) = \tilde{W}$ for all process matrices $W$. The process $\tilde{W}$ can have any causal structure. One can also define a map that trivially interpolates between $W$ and $\tilde{W}$ for any process $W$, that is, a map $\A$ such that $\A(W) = (1-p) W + p\tilde{W}$, for $p$ continuously varying between 0 and 1. This map leaves the input $W$ unchanged with probability $1-p$ and prepares the process $\tilde{W}$ with probability $p$. It is an example of a continuous (yet non reversible) transformation (more precisely, continuously connected to the identity) that changes the causal order. 

Although the formalism for process matrix transformations allows to change the causal structure of a process trivially, the examples presented rely implicitly on the physical realisability of the process $\tilde{W}$, which might not be clear to begin with. Therefore, it is natural to require that the transformation $\A$     
satisfy reasonable physicality conditions in addition to conditions (\ref{Aconditions}).

\subsection{$W_{OCB}$ cannot be obtained from a causally separable process via a reversible transformation}
In view of the fact that process matrix transformations with no restrictions other than the consistency conditions (\ref{Aconditions}) can trivially produce an output process matrix that has a different causal structure to the initial process matrix, we now consider reversible transformations. This is a natural requirement on transformations of states in all Generalised Probabilistic Theories (GPT) approaches to reconstructions of Quantum Theory \cite{hardyreconstruction, borireconstruction, muellerreconstruction}. As states are a particular subclass of process matrices, it is natural to assume the same requirement for a general process matrix.  

In particular, we consider reversible, but not necessarily continuous, transformations between bipartite process matrices with two-dimensional input and output Hilbert spaces. Let us consider the process matrix $W_{OCB}$ as defined in Eq. (\ref{wocb}) and ask whether there exists a reversible process matrix transformation that reaches $W_{OCB}$ from a causally order process matrix. As we will show now, there is no reversible process matrix transformation that can achieve such a task, suggesting that $W_{OCB}$ is not physically realisable.  

The first thing we note is that, as shown in Appendix (\ref{appendix2}), $W_{OCB}$ is an extremal process, in the sense that $W_{OCB} = \lambda W^\prime_1 + (1-\lambda)W^\prime_2$ implies $W^\prime_1 = W^\prime_2$. If $\A(W) = W_{OCB}$, then $W$ has to be extremal, since $W = \mu W_1 + (1-\mu) W_2$ implies $W_{OCB} = \mu \A(W_1) + (1-\mu )\A(W_2)$, which means $\A(W_1) = \A(W_2)$. Since $\A$ is an injective map, we have $W_1 = W_2$, showing that $W$ is extremal. If we require that $W$ be causally separable, the fact that it is extremal implies that it is causally ordered. 

The second thing we note is that $W_{OCB}$ is a rank-8 process matrix. Because reversible transformations are rank-preserving, it follows that $W$ has to be an extremal, causally ordered, rank-8 process. We will show that no such process exists. Without loss of generality, assume that $W = W^{A\leq B}$ is a channel (possibly with memory) from $A$ to $B$. The following theorem -- proven by D'Ariano et al. \cite{dariano} and adapted here to bipartite causally ordered W matrices with signalling from $A$ to $B$ -- provides necessary and sufficient conditions for $W^{A\leq B}$ to be extremal: 
\textit{Let $W^{A\leq B}$ be a causally ordered process matrix and let $\mathrm{supp}( W^{A \leq B})$ be its support. Let $C$ be a basis of hermitian operators in the space $\mathcal{L}(\mathrm{supp}(W^{A\leq B}))$. Define $D = \{\sigma^{B_I}_i\sigma^{B_O}_\mu\sigma^{A_I}_\nu\sigma^{A_O}_\rho, \mathbb{1}^{B_I}\mathbb{1}^{B_O}\sigma^{A_I}_i\sigma^{A_O}_\mu \}$, for greek indices running from 0 to 3 and latin indices from 1 to 3. The process $W^{A\leq B}$ is extremal iff the disjoint union $C \sqcup D$ is linearly independent.} 	

From this theorem it is straightforward to see that there is no extremal, rank-8, causally ordered bipartite process. The process being rank 8 implies that $C$ has $8^2 = 64$ elements. The set $D$ has $204$ elements, which means their disjoint union $C \sqcup D$ has $268$ elements. The dimension of $\mathcal{L(\mathcal{H})}$ is $256$, implying that $C \sqcup D$ is not linearly independent. This shows that $W^{A\leq B}$ is not extremal. We conclude that $W_{OCB}$ cannot be obtained reversibly from a causally ordered process. Note that the restrictions we impose on the transformations in this example are weaker than those leading to our result in \ref{continuousandreversible}, that is, we ask for the transformations to be reversible but not necessarily continuous. Nevertheless, as we have shown, the process (\ref{wocb}) cannot be obtained by transforming a causally ordered one, even without the continuity restriction. 

\subsection{C-SWAP transformation: obtaining the quantum switch from a channel}
We have shown that the process matrix (\ref{wocb}) cannot be obtained from a causally ordered process via a reversible transformation. We will now show that this is not the case for the quantum switch: There exists a reversible process matrix transformation that takes a causally ordered process (a channel in one direction) to the quantum switch. 

Consider the Hilbert space corresponding to the quantum switch $\mathcal{H} = \mathcal{H}^{A_I}\otimes\mathcal{H}^{B_I}\otimes\mathcal{H}^{A_O}\otimes\mathcal{H}^{B_O}\otimes\mathcal{H}^{C_T}\otimes\mathcal{H}^{C_C}$ and the transformation 
\begin{equation}
V = \mathbb{1}^{A \, B \, C_T}\otimes\ketbra{0}{0}^{C_C} + SWAP^{A \,B} \otimes \mathbb{1}^{C_T}\otimes\ketbra{1}{1}^{C_C},	
\end{equation}  
where for simplicity of notation we have written $A$ to denote the Hilbert space $\H^A = \H^{A_I}\otimes\H^{A_O}$ and similarly for $B$. We define $SWAP^{A \,B}  = SWAP^{A_I \,B_I}\otimes SWAP^{A_O \,B_O} $ where the well-known $SWAP$ operator acts as $SWAP \ket{\psi}\otimes\ket{\phi} = \ket{\phi}\otimes\ket{\psi}$. The transformation $V$ is a proper process matrix transformation in the sense that it obeys the conditions (\ref{Aconditions}), and yields the quantum switch when acting on the channel $\ket{ABC}^{ABC_T}\ket{+}^{C_C}$, with the notation used in (\ref{switch}) and $\ket{+} = (\ket{0}+\ket{1})\sqrt{2}$:
\begin{equation}
V \ket{ABC}^{ABC_T}\ket{+}^{C_C}  = \frac{1}{\sqrt{2}}\ket{ABC}^{ABC_T}\ket{0}^{C_C} + \frac{1}{\sqrt{2}}\ket{BAC}^{ABC_T}\ket{1}^{C_C}  = \ket{S}.	
\end{equation}

The transformation $V$ is a reversible supermap that takes a causally separable process as an input (a quantum channel) to a causally non-separable process (the quantum switch). One could ask if this transformation can be cast in such a way that it is also continuous, i.e. if there exists a continuously parametrised set of supermaps that reduces to $V$ for some value of the parameter and to the identity for some other value. A seemingly good candidate is the transformation $V_\lambda = \mathrm{exp}(-\mathrm{i}\lambda H)$, for $H = (SWAP^{A_I \,B_I}\otimes SWAP^{A_O \,B_O} - \mathbb{1}^{A_I \,B_I}\otimes \mathbb{1}^{A_O \,B_O})\otimes\mathbb{1}^{C_T} \otimes \ketbra{1}{1}^{C_C}$. This transformation reduces to the identity for $\lambda = 0$ and to $V$ for $\lambda = \pi /2$. However, applying $V_\lambda$ to the process  vector $\ket{ABC}^{ABC_T}\ket{1}^{C_C}$ yields $V_\lambda \ket{ABC}^{ABC_T}\ket{1}^{C_C} = \mathrm{exp}(\mathrm{i}\lambda)(\cos\lambda\ket{ABC}^{ABC_T}-\mathrm{i}\sin\lambda\ket{BAC}^{ABC_T})\ket{1}^{C_C}$, which is not a valid process matrix since, as a straightforward calculation shows, it contains, for instance, ``forbidden'' terms of the form $\sigma_i\otimes\sigma_i\otimes\sigma_i\otimes\sigma_i\otimes\mathbb{1} \in \mathcal{L}(\mathcal{H}^{A_I}\otimes\mathcal{H}^{B_I}\otimes\mathcal{H}^{A_O}\otimes\mathcal{H}^{B_O}\otimes\mathcal{H}^{C_I})$. The failure to find a continuous and reversible process matrix transformation that changes the causal structure is a particular instance of the more general result of Section (\ref{continuousandreversible}), namely, that all continuous and reversible transformations of process matrices reduce only to local unitaries in the parties' input and output Hilbert spaces, showing that no continuous and reversible process matrix transformation can change the causal structure.

\begin{figure}[h]

\centering
\includegraphics[scale=0.25]{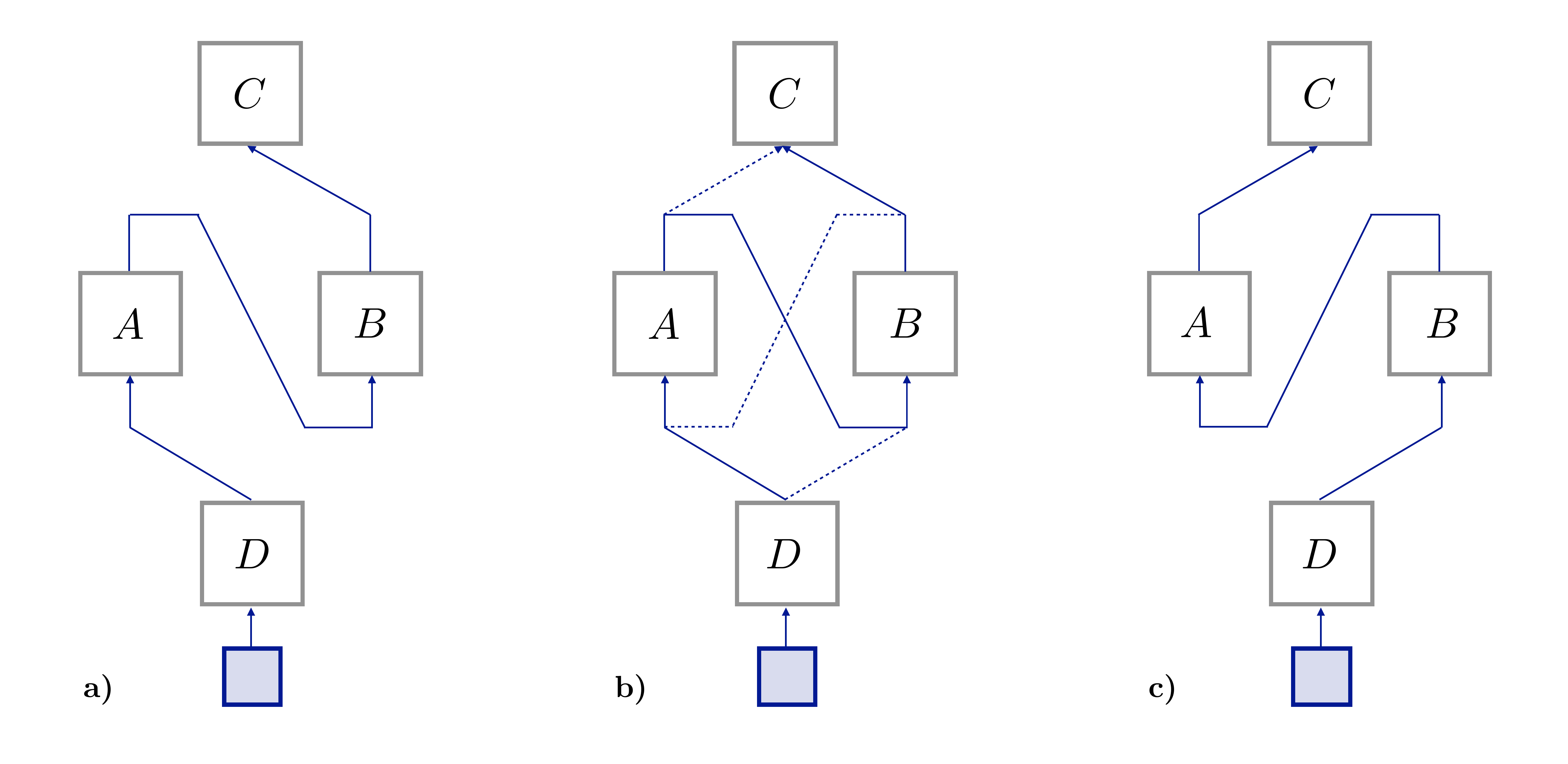}
\caption{\label{extendedswitch}A four partite process matrix for parties $A$, $B$, $C$ and $D$ can be mapped to a continuous family of three partite processes for parties $A$, $B$ and $C$. By choosing different local operations for $D$, one can map the four partite input process matrix to processes that exhibit different causal order. In a concrete quantum optical realisation of this transformation, $D$ controls the reflectivity of a beam splitter (see main text). The transformation can go continuously from \textbf{a)} a channel from $A$ to $B$ to $C$ to \textbf{c)} a channel from $B$ to $A$ to $C$, passing through \textbf{c)} the quantum switch (see Eq. (\ref{switch})).}
\end{figure}

\subsection{Local operations of one party that influence the causal structure of the remaining parties}
It is known \cite{oreshkov, abbot} that, in a general multipartite process matrix, local operations performed by one of the parties can influence the causal structure of the remaining parties. We will show now that this feature can be expressed naturally in terms of process matrix transformations. Consider an $N$ partite process $W_N$ over a Hilbert space $\H_N$ for $N>2$ and let the $X$-th party perform a local $CPTP$ map $\M$ in his/her laboratory. This will induce an $N-1$ \textit{reduced} process matrix $W_{N-1}$ over the Hilbert space $\H_{N-1}$. The causal order of the reduced process can in general depend on $\M$. Define the process matrix transformation $\A_\M: \mathcal{L}(\H_N)\longrightarrow \mathcal{L}(\H_{N-1})$ as 
\begin{equation}
\A_\M (W) = Tr_X \left(C_\M W\right),
\end{equation}
where $C_\M$ denotes the CJ representation of the map $\M$. The supermap $\A_\M$ is a well-defined process matrix transformation that can in principle yield processes with indefinite causal structure. Note, however, that the process $W_{N-1}$ being causally non separable implies that the process $W_N$ itself is also causally non separable. To see this, suppose that $W_N$ is causally separable. This means that for \textit{any} choice of local operations, the resulting multipartite experiment can always be explained in terms of a convex combination of states and channels. This fact holds for the particular case in which the $X$-th party chooses to apply the map $\M$. But this choice would imply that the process $W_{N-1}$ is causally separable, contradicting our initial assumption. 

Let us now analyse a concrete example of a process matrix transformation where the action of a party in the past decides the direction of the signalling of parties in the future. This party can even decide whether the remaining process is causally separable or not. Our example, depicted in Fig. (\ref{extendedswitch}) is motivated by the experimental implementation of the quantum switch in an optical setup \cite{procopio, giulia}. In this experiment, a photon traverses a beam splitter that divides its trajectory into two paths. In one path the local operation $\M_A$ is applied to the polarisation degree of freedom of the photon before local operation $\M_B$, while the other path corresponds to exactly the opposite: The operation $\M_B$ is followed by $\M_A$. By continuously changing the reflectivity of the beamsplitter, we can go continuously from a situation in which $A$ signals to $B$ to the situation in which $B$ signals to $A$, passing through an intermediate situation the corresponds to the quantum switch. 

Formally, the action of changing the reflectivity of the beamsplitter corresponds to adding an extra party $D$ to the quantum switch (\ref{switch}). The initial process $W$ that we consider has therefore a Hilbert space $\H =\mathcal{H}^{A_I}\otimes\mathcal{H}^{B_I}\otimes\H^{D_I}\otimes\mathcal{H}^{A_O}\otimes\mathcal{H}^{B_O}\otimes\H^{D_O}\otimes\mathcal{H}^{C_T}\otimes\H^{C_C}$. $W$ is given by $W = \ketbra{W}{W}$, where
\begin{equation}
\ket{W} = \ket{ABC}^{ABC_T}\ket{0}^{D_I}\ket{0}^{D_O}\ket{0}^{C_C} + \ket{BAC}^{ABC_T}\ket{0}^{D_I}\ket{1}^{D_O}\ket{1}^{C_C},	
\end{equation}
using the same notation as in Eq. (\ref{switch}). We consider the case in which $D$ performs a unitary map $\U: \H^{D_I}\longrightarrow \H^{D_O}$ defined as $\U(\rho) = U \rho U^\dagger$, specified by the matrix
\begin{equation}
U=
\begin{pmatrix}
	\cos\lambda && -\sin\lambda\\
	\sin\lambda && \cos\lambda	
\end{pmatrix}.
\end{equation}     	
The output process is obtained when $D$ acts locally with $\U$. This defines a supermap $\A_\U$ in the sense described above. Its action on $W$ yields $\A_\U(W) = \mathrm{Tr}_{D}\left(C_\U W\right) = \ketbra{W_\lambda}{W_\lambda}$, where
\begin{equation}
\ket{W_\lambda} = \cos\lambda\ket{ABC}^{ABC_T}\ket{0}^{C_C}+\sin\lambda\ket{BAC}^{ABC_T}\ket{1}^{C_C}.	
\end{equation} 
$\ket{W_\lambda}$ is a channel from $A$ to $B$ for $\lambda =0$, reduces to the quantum switch for $\lambda =\pi/2$, and to a channel from $B$ to $A$ for $\lambda =\pi$. The party $D$ can change the causal structure of the remaining three parties by applying a transformation locally. Note that, although the transformation applied by $D$ is continuous and reversible, the map $\A_\U$ is not reversible, because it sends a four-partite process matrix into a three-partite one. Note that the necessity for adding an extra party $D$ to describe the change in reflectivity of the beamsplitter comes from the definition of process matrices as operations that yield objects with no open ends. Because process matrices are conveniently \textit{defined} to describe the physics of everything external to the local laboratories, had we not added the party $D$, the probabilities and the causal order of operations would have not been determined only by the local operations and the wiring between them, but also by the ``external'' system operating on the beamsplitter. In fact, as we showed above, no continuous and reversible process matrix transformation can change the causal structure.

\section{discussion}
\label{discussion}

The process matrix formalism is an operational quantum framework that assumes the existence of a definite causal order between events \textit{locally}, while making no assumption regarding the \textit{global} causal structure. This resembles the situation of general relativity, where one arrives at the notion of a curved space-time from the assumption that special relativity holds locally.

Process matrices reduce to a state or to the time evolution of a state (channel) when the global causal order between events is definite, but represent a more general notion when the causal order is indefinite. The latter may appear, for example, when matter degrees of freedom of quantum systems are prepared in a state of spatial superposition, such that the metric induced and consequently the spatiotemporal distances between events become indefinite \cite{magdalena, feix}. 

In this paper we have developed a theory of dynamics of process matrices, or equivalently, transformations of process matrices into process matrices. Just like states and channels, general process matrices can be dynamical. For example, if we consider an arrangement of laboratories, together with the corresponding physical systems traversing them, occupying a region of space-time, the process matrix describing the experiments carried out in the laboratories can be subjected to transformations as a consequence of the different values of the metric field along the space-time trajectory of the arrangement. 

We have focused on transformations that are \textit{continuous and reversible}. This restriction is motivated by the fact that the evolution of physical systems that we know in nature are of this type. In fact, continuous and reversible transformations of physical states have been considered as axioms from which one derives quantum theory \cite{hardyreconstruction, borireconstruction, muellerreconstruction}. We have shown that no continuous and reversible transformation can change the causal order of a process, because all such transformations amount to local unitary operations in the parties' input and output spaces. Crucially, in order to apply our result, the process under consideration has to yield a ``closed system'', meaning that all the probabilities predicted by the process are specified by $(\mathrm{1})$: The local operations performed by the parties and $(\mathrm{2})$: The way the local laboratories are ``connected''. For the causally ordered case, this connection is specified by all the ``wires'' that link the different local laboratories. For the general case, where causality can be indefinite, the wires are exchanged by an ``E-shaped'' diagram representing a process matrix. Once the process yields a closed system in the specified sense, its causal order cannot change if we subject it to continuous and reversible transformations. This fact shows that there are strong restrictions on the physical transformations that might be needed in order to implement processes with indefinite causal structure. Under the assumption that that all transformations of processes are continuous and reversible, our results might provide an explanation as to why no process that violates causal inequalities has been realised physically by ``evolving'' causally ordered processes.

The type of dynamics of causal structures introduced here requires transformations that output valid process matrices for \textit{any} input process matrix on which the transformation acts. However, one could think of a different type of restricted transformations that are well defined only for certain classes of process matrices, e.g. processes having the same causal structure (or, for example, transformations that apply only for the case of states or only for the case of channels). Although such transformations would not capture the dynamics of causal structures in a unified fashion, they might prove to be relevant because of the less stringent conditions imposed on them.

Our formalism captures the notion that, in a multipartite process, local operations of one of the parties can alter the causal order of the remaining parties, as noted in \cite{oreshkov, abbot}. In these cases, it is important to note that if one can obtain a causally indefinite process via the action of a local operation by one of the parties, then the total process, including this party, is causally non separable to begin with. 

As a final remark, we note that, as for the case of quantum states, any process matrix transformation has a dilation, that is, a representation in terms of a unitary operator acting on the joint Hilbert space of the process matrix and an ancillary system. In contrast to the case of quantum states, however, it is not guaranteed that the initial process matrix remains a valid process throughout its unitary evolution coupled to the ancilla, before tracing it out. One might call a dilation ``physical'' if it preserves the validity of all process matrices throughout the unitary evolution, before tracing the ancilla out. Given the important role that dilations play for the realisation of quantum networks \cite{chiribellacombs, bisio1, bisio2}, it is an interesting question whether causally separable processes can ``evolve'' into causally non separable ones via transformations that have only continuous, physical dilations. In view of the strong restrictions on continuous unitary transformations found in this work, it seems reasonable to conjecture a negative answer to this question. A rigorous proof of this fact is left for future work.

\section{Acknowledgements}
We thank P. Allard-Gu\'erin, M. Ara\'ujo, C. Budroni, F. Costa, P. Perinotti and M. Sedl\'ak for interesting discussions. We acknowledge the support from the Austrian Science Fund (FWF) through the Special Research Programme FoQuS, the Doctoral Programme CoQuS and the project I-2526 and the research platform TURIS. This publication was made possible through the support of a grant from the John Templeton Foundation. The opinions expressed in this publication are those of the authors and do not necessarily reflect the views of the John Templeton Foundation.

\onecolumngrid
\newpage
\appendix

\section{Further details on the characterisation of process matrices}
\label{appendixN1}
In this section we give, for the sake of completeness, a description of the projector $P$ introduced on Section \ref{processmatrices} in the main text. For our purposes, it suffices to focus on the bipartite case, corresponding to the Hilbert space $\H = \H^{A_I} \otimes \H^{B_I} \otimes \H^{A_O} \otimes \H^{B_O}$. Let the two parties, $A$ and $B$, perform local operations described by completely positive (CP) maps $\M_i^{X}$, $X = A, B$. The label $i$ denotes a possible measurement outcome of the corresponding local operation. The CP maps act from the input Hilbert space $\H^{X_I}$ to the output Hilbert space $\H^{X_O}$ of each party $X = A, B$. The set of local operations $\{\M_i^{X}\}$, where $i$ takes values on the set of possible outcomes, form a quantum instrument, that is, they add up to a completely positive, trace preserving (CPTP) map $\M^X = \sum_i\M_i^{X}$. The joint probability for $A$ to obtain outcome $i$, corresponding to the operation $\M_i^{A}$, and for $B$ to obtain outcome $j$, corresponding to the operation $\M_j^{B}$, is given by the ``generalised Born rule'' (Eq. (\ref{brule})): $p_{ij} = \Tr{}\left(W C_{\M^A_{i}} \otimes C_{\M^B_{j}}\right)$, where $W \in \mathcal{L}(\H)$ is the process matrix describing the physics outside the local laboratories, and $C_{\M^X_{i}}$ denotes the Choi-Jamio\l kowski (CJ) representation of the map $\M^X_{k}$, for $X = A, B$ and $k = i,j$. The conservation of probability means that the sum of all probabilities equals unity: $\sum_{ij} p_{ij} = \Tr{}\left(W C_{\M^A} \otimes C_{\M^B}\right) = 1$. It is known that the CJ representation of a map satisfies $\Tr{X_O} C_{\M^X} = \mathbb{1}^{X_I}$ if and only if the map $\M^X$, $X = A, B$, is trace preserving. Therefore, the conservation of probability for process matrices can be stated as 
\begin{equation}
\Tr{}\left(W C_{\M^A} \otimes C_{\M^B}\right) = 1,
\end{equation}
for all $C_{\M^X}$, $X = A, B$, satisfying $\Tr{X_O} C_{\M^X} = \mathbb{1}^{X_I}$. It was shown in Refs. \cite{ocb, witness}, that this condition implies, apart from the normalisation of the trace of $W$ given in Eq. (\ref{wcondition2}), $\Tr{}\left(W\right) = d_{O}$, that the process matrix $W$ belongs to a proper subspace of the total Hilbert space of operators $\mathcal{L}(\H)$. Explicitly, $W$ is invariant under the projector $P$ introduced in Ref. \cite{witness}, given by
\begin{equation}
\label{projectionop}
P(W) = _{A_O}\!\!W+_{B_0}\!\!W-_{A_I A_O}\!\!W-_{B_I B_O}\!\!W-_{A_O B_O}\!\!W + _{A_O B_I B_O}\!\!W + _{B_O A_I A_O}\!\!W,	
\end{equation}
where $_X\!W := d_X^{-1}\mathbb{1}^X\otimes \Tr{X}(W)$ for any subspace $X$ of $\mathcal{L}(\H)$.

In order to acquire some physical intuition regarding the structure of $P$, it is useful to write down the elements of $\mathcal{L}(\H)$ in terms of the Hilbert-Schmidt basis $\mathrm{B} = \{\sigma_\alpha \otimes \sigma_\beta \otimes \sigma_\gamma \otimes \sigma_\delta\}$, where all greek indices run from 0 to $d^2-1$, with $d$ equal to the dimension of a single tensor factor of the total space, for example $\H^{A_I}$. For simplicity of notation, we assume that each tensor factor of $\mathcal{H}$ is equal to the space $\mathbb{C}^d$. Note that this assumption can be made without losing generality, because we can always add dimensions to laboratories with trivial operations. By definition, $\sigma_0 = \sqrt{2/d} \,\mathbb{1}$, and $\sigma_i$ is a hermitian, traceless operator for all values of the latin index $i$, running from 1 to $d^2-1$. In this basis, the projector defined in Eq. (\ref{projectionop}) has a particularly simple form. Note that the operator $\sigma_0\otimes\sigma_0\otimes\sigma_0\otimes\sigma_0$ is trivially left invariant by $P$. The other elements of $\mathrm{B}$ invariant under $P$ are called ``valid'' or ``allowed'' terms. The condition given by Eq.(\ref{wcondition3}) then implies that any process matrix can be written as a (positive and suitably normalised) linear combination of the identity operator and valid terms.

Following \cite{ocb}, we briefly list the set of different valid terms and mention their physical meaning. Terms of the form  $\sigma_\alpha \otimes \sigma_i \otimes \sigma_j \otimes \sigma_0$ describe channels without memory (for $\alpha = 0$) and channels with memory (for $\alpha\neq0$) from $A$ to $B$ \cite{ocb}. A process matrix containing a term of this form describes signaling correlations from $A$ to $B$. Analogously, terms of the form $\sigma_i \otimes \sigma_\beta \otimes \sigma_0 \otimes \sigma_j$ represent channels, possibly with memory, from $B$ to $A$. Terms of the form $\sigma_\alpha \otimes \sigma_\beta \otimes \sigma_0 \otimes \sigma_0$ represent states shared between $A$ and $B$. As the states might be entangled, all non-signaling correlations are described by process matrices containing these terms.

It is also instructive to look at the so-called ``invalid'' or ``forbidden'' terms $F$, that is, terms for which $P(F) = 0$. These terms are always absent in any valid process matrix since, as we will discuss with some examples, their presence would imply that probabilities are not conserved. Consider first terms of the form $\sigma_0 \otimes \sigma_0 \otimes \sigma_i \otimes \sigma_\alpha$ and $\sigma_0 \otimes \sigma_0 \otimes \sigma_\alpha \otimes \sigma_i$. These terms can be interpreted, respectively, as post-selection of measurement results for party $A$, and post-selection of measurement results for party $B$. When $\alpha\neq 0$, both parties perform post-selection. Clearly, post-selection terms lead to the non-conservation of probability: If, for example, we choose quantum instruments that re-prepare the system in a state which is orthogonal to the post-selection subspace, then the post-selected state is never realised and all probabilities will vanish. 

Let us now examine terms of the form $\sigma_i \otimes \sigma_0 \otimes \sigma_j \otimes \sigma_\alpha$ and $\sigma_0 \otimes \sigma_i \otimes \sigma_\alpha \otimes \sigma_j$. These terms can be interpreted, respectively, as ``local loops'' in $A$ and ``local loops'' in $B$. When $\alpha\neq0$, these terms involve also post-selection. Local loops describe signaling from the output of a party's laboratory to the input of the same party. As noted in Section \ref{processmatrices}, such terms correspond to closed time-like curves. They allow a party to send a signal into her/his past and give rise to ``grandfather-type'' paradoxes. Processes containing local loops do not conserve probabilities. As a simple example of this fact, consider a case where only $A$ is non trivial, and let her input and output spaces be two-dimensional. The total Hilbert space is then $\H^A = \H^{A_I} \otimes \H^{A_O} \approx \mathbb{C}^2\otimes \mathbb{C}^2$. Let the ``process'' be given by $W_{LL} = \dketbra{\mathbb{1}}{\mathbb{1}}$. This process corresponds to an identity channel from the output of $A$ to the input of $A$. Note that $W_{LL}$ satisfies all conditions for process matrices except the one given by Eq.(\ref{wcondition3}). Now consider the quantum instrument $\{M_0 = \ketbra{0}{0}\otimes\ketbra{1}{1}, M_1 = \ketbra{1}{1}\otimes\ketbra{0}{0}\}$, describing a projective measurement of the system in the basis $\{\ket{0},\ket{1}\}$ and a re-preparation in the state $\ket{1}$, for the outcome 0 and in the state $\ket{0}$, for the outcome 1. Note that this scenario leads to a paradoxical situation in which, loosely speaking, ``0 equals 1''. This is just an instance of the grandfather paradox. Using Eq.(\ref{brule}) to calculate the probability $p_i$ to obtain outcome $i$, $i = 0,1$, we find that both probabilities vanish, violating conservation of probability. Finally, terms of the form $\sigma_i \otimes \sigma_j \otimes \sigma_k \otimes \sigma_l$ are called ``global loops''. These terms also lead to paradoxical situations and non conservation of probability. In particular, if one of the parties performs an identity channel as his/her local operation, a global loop becomes effectively a local loop and leads to the consequences discussed above.       
 
\section{Complete positivity of process matrix transformations}
\label{appendixN2}
Let $\mathcal{A}:\mathcal{L}(\mathcal{H}_1)\longrightarrow\mathcal{L}(\mathcal{H}_2)$ be a valid process matrix transformation. In general, $\mathcal{L}(\mathcal{H}_1)$ and $\mathcal{L}(\mathcal{H}_2)$ are tensor products of Hilbert spaces corresponding to many parties, each of them with input and output subspaces. If $W_1 \in \mathcal{L}(\H_1)$ is a valid process matrix, then $\A (W_1) = \Tr{1}(C_\A W_1^T\otimes\mathbb{1}) \in \mathcal{L}(\mathcal{H}_2)$ is a valid process matrix. Because it will be useful below, we have expressed the action of $\A$ on $W_1$ in terms of the inverse of the Choi-Jamio\l kowski isomorphism. As in the main text, $C_\A \in \mathcal{L}(\H_1\otimes\H_2)$ denotes the CJ representation of $\A$. Now suppose there exists an additional Hilbert space $\mathcal{H}^\prime_1$ such that $\mathcal{L}(\mathcal{H}^\prime_1)$ also possesses an input-output structure with two or more parties. Now let $W\in \mathcal{L}(\mathcal{H}_1\otimes\mathcal{H}^\prime_1)$ be a be a valid process matrix. The map $\A\otimes\mathbb{1}:\mathcal{L}(\mathcal{H}_1\otimes\mathcal{H}^\prime_1)\longrightarrow\mathcal{L}(\mathcal{H}_2\otimes\mathcal{H}^\prime_1)$ is a transformation that acts only on a subset of the parties that compose $W$, namely, those corresponding to the space $\mathcal{L}(\mathcal{H}_1)$, leaving the others intact. Clearly, this transformation is physically sound, and we must therefore demand that it is a valid process matrix transformation for all (finite dimensional) Hilbert spaces $\mathcal{H}^\prime_1$. We will show that this requirement, together with the assumption that the parties of the transformed process matrix can share entangled ancillary systems (as discussed, for example, in \cite{witness}), implies that $\A$ is a completely positive map. For concreteness, we assume that $\A$ transforms bipartite process matrices into bipartite process matrices. The extension to more general situations is straightforward.

The complete positivity of $\A$ is equivalent to the non-negativity of its Choi-Jamio\l kowski (CJ) representation, $C_\A$. In what follows we show that, under the assumptions stated in the previous paragraph, $C_\A$ is indeed non-negative. The action of $\A\otimes\mathbb{1}$ on $W$ can also be written in terms of $C_\A$ by means of the inverse of the Choi-Jamio\l kowski isomorphism: $\A\otimes\mathbb{1} (W) = \Tr{1}(C_\A W^{T_1}\otimes\mathbb{1})$, where the subscript $1$ denotes the ``initial'' Hilbert space $\H_1$, the superscript $T_1$ denotes transposition with respect to $\mathcal{H}_1$, and the identity operator acts on $\H_2$. Let $W\in \mathcal{L}(\mathcal{H}_1\otimes\mathcal{H}^\prime_1)$ be a four-partite process matrix with two parties, $A$ and $B$ corresponding to the Hilbert space $\H_1 = {H}^{A_I}\otimes\mathcal{H}^{B_I}\otimes\mathcal{H}^{A_O}\otimes\mathcal{H}^{B_O}$, and two parties $A^\prime$ and $B^\prime$ corresponding to the Hilbert space $\H^\prime_1 = {H}^{A^\prime_I}\otimes\mathcal{H}^{B^\prime_I}\otimes\mathcal{H}^{A^\prime_O}\otimes\mathcal{H}^{B^\prime_O}$. For our purposes, we can consider the output space of $\H^\prime_1$ to be trivial and refer to $A^\prime_I$ and $B^\prime_I$ simply as $A^\prime$ and $B^\prime$. The transformation $\A$ has a CJ representation $C_\A \in \mathcal{L}(\mathcal{H}_1\otimes\mathcal{H}_2)$, where $\mathcal{H}_2 = {H}^{C_I}\otimes\mathcal{H}^{D_I}\otimes\mathcal{H}^{C_O}\otimes\mathcal{H}^{D_O}$ corresponds to the ``final'' parties $C$ and $D$. We assume that the four parties of the  ``final'' process, $A^\prime$, $B^\prime$, $C$, $D$, share a (possibly entangled) ancillary state. For this purpose, let $\rho^T$ be an arbitrary state in a supplementary Hilbert space $\H^{\prime\prime} = \H^{A^{\prime\prime}} \otimes \H^{B^{\prime\prime}} \otimes \H^{C^{\prime\prime}} \otimes \H^{D^{\prime\prime}}$ and consider the laboratories corresponding to the labels $A^\prime$ and $ A^{\prime\prime}$ as a single laboratory, as well as the laboratories corresponding to the labels $B^\prime$ and $B^{\prime\prime}$, $C$ and $C^{\prime\prime}$, and $D$ and $D^{\prime\prime}$. The non-negativity of probabilities then reads
\begin{equation}
\label{positivitycond}
0 \leq \Tr{}\left(C_\A^{ABCD}(W^{T_{AB}})^{ABA^\prime B^\prime} E^{A^\prime A^{\prime \prime}} E^{B^\prime B^{\prime \prime}} M^{C C^{\prime \prime}} M^{D D^{\prime \prime}}(\rho^T)^{A^{\prime \prime}B^{\prime \prime}C^{\prime \prime}D^{\prime \prime}}\right),	
\end{equation}
where $E^{A^\prime A^{\prime \prime}}$ and $E^{B^\prime B^{\prime \prime}}$ denote, respectively, positive operator valued measure (POVM) elements in the spaces of $A^\prime$ and $A^{\prime\prime}$ and of $B^\prime$ and $B^{\prime\prime}$, and $M^{C C^{\prime \prime}} $ and $ M^{D D^{\prime \prime}}$ denote, respectively, the CJ representations of CP maps from the input spaces to the output spaces of the parties $C$ and $C^{\prime \prime}$ and of $D$ and $D^{\prime \prime}$. In Eq. (\ref{positivitycond}) above, we have adopted a notation in which the super-indices label the Hilbert spaces in which the matrices act. For example, the label $A$ denotes the Hilbert space $\H^A = \H^{A_I}\otimes\H^{A_O}$. The superscript $T_{AB}$ denotes the partial transpose with respect to $\H^A\otimes\H^B$. 

In order to prove the non-negativity of $C_\A$, we consider the case in which the following Hilbert space isomorphisms hold: $\H^{X^{\prime\prime}} = \H^{X_1^{\prime\prime}}\otimes \H^{X_2^{\prime\prime}} \approx \H^{X^\prime} = \H^{X_1^{\prime}}\otimes \H^{X_2^{\prime}} \approx \H^X = \H^{X_I}\otimes \H^{X_O}$, for $X = A,B$; and $\H^{X^{\prime\prime}} = \H^{X_1^{\prime\prime}}\otimes \H^{X_2^{\prime\prime}} \approx \H^X = \H^{X_I}\otimes \H^{X_O}$, for $X = C,D$. Here we have defined the subspaces $\H^{X_i^{\prime\prime}}$, for $X = A, \ B, \ C, \ D$ and $i = 1,\ 2$, and the subspaces $\H^{X_i^{\prime}}$, for $X = A, \ B$ and $i = 1,\ 2$ in order to give an explicit tensor product structure to the primed and double-primed Hilbert spaces. We now choose the operations $E^{X^\prime X^{\prime \prime}} = \ketbra{\Phi^+}{\Phi^+}^{X^\prime X^{\prime \prime}}$, where $\ket{\Phi^+}$ denotes the (normalised) maximally entangled state, for $X = A,B$, and  $M^{X X^{\prime \prime}} = \dketbra{\mathbb{1}}{\mathbb{1}}^{X_1^{\prime \prime}X_I}\otimes \ketbra{\Phi^+}{\Phi^+}^{{X_2^{\prime \prime}X_O}}$, for $X = C,D$. Let the process matrix be given by $W^{ABA^{\prime} B^{\prime}} = \dketbra{\mathbb{1}}{\mathbb{1}}^{AA^\prime}\otimes\dketbra{\mathbb{1}}{\mathbb{1}}^{BB^\prime}$. Note that $W$ is a valid process matrix that represents channels with memory from $A$ to $A^\prime$ and from  $B$ to $B^\prime$. It is straightforward to check that this choice of operations and process matrix, when inserted in Eq. (\ref{positivitycond}), lead to
\begin{equation}
0 \leq \Tr{}(C_\A^{ABCD} \rho^{ABCD}).	
\end{equation}
Since $\rho$ is an arbitrary (normalised) positive matrix, the non-negativity of $C_\A$ follows.

\section{Characterisation of continuous and reversible process matrix transformations}
\label{appendix1}
As mentioned in Section \ref{continuousandreversible} of the main text, a continuous, reversible transformation $\mathcal{U}$ from process matrices to process matrices is of the form $\mathcal{U} (W) = U W U^\dagger$, for a unitary operator $U = \mathrm{e}^{\mathrm{i}\lambda H}$. The hermitian and traceless operator $H$ satisfies 
\begin{equation}
\label{conditionsup}
P(\left[H,W\right]) = \left[H,W\right] 
\end{equation}
for all valid processes $W$. In the following, we show that every transformation of this form consists only of local unitary operations. That is, we show that $H$ contains only single-body terms. For simplicity, we focus on the bipartite case; the proof can then be generalised straightforwardly to an arbitrary number of parties. 

Let $\mathcal{H} = \mathcal{H}^{A_I}\otimes\mathcal{H}^{B_I}\otimes\mathcal{H}^{A_O}\otimes\mathcal{H}^{B_O}$ be the Hilbert space on which bipartite process matrices act. As in Appendix A, we assume that each tensor factor of $\mathcal{H}$ is equal to the space $\mathbb{C}^d$, without loss of generality. In terms of the Hilbert-Schmidt basis, $H$ can be written as
\begin{equation}
H = h_{\alpha \beta \gamma \delta} \, \sigma_\alpha \otimes \sigma_\beta \otimes \sigma_\gamma \otimes \sigma_\delta.	
\end{equation}
Here, as in Appendix A, $\sigma_0 = \sqrt{2/d} \,\mathbb{1}$, and the operators $\sigma_i$, with $1\leq i \leq d^2-1$, are the (traceless and hermitian) generators of the Lie algebra $\mathrm{su}$$(d)$. They satisfy
\begin{equation}
\label{basic}
\sigma_i\sigma_j = \delta_{ij}\mathbb{1} + \mathrm{d}_{ijk}\sigma_k + \mathrm{i}\mathrm{f}_{ijk}\sigma_k,
\end{equation}
where $\mathrm{d}_{ijk}$ is totally symmetric and traceless (i.e. $\mathrm{d}_{iik} = 0$), and the coefficients $\mathrm{f}_{ijk}$, called the structure constants of $\mathrm{su}$$(d)$, are totally anti-symmetric (for further details, consult Ref. \cite{bengtsson}). In our notation, greek indices run from 0 to $d^2-1$, and latin ones from 1 to $d^2-1$. When an index is repeated, we assume that it is summed over.

Requiring condition (\ref{conditionsup}) is equivalent to requiring 
\begin{equation}
\label{eqcondition}
\mathrm{Tr}\left(\left[H,T\right]F\right) = 0	
\end{equation}
for all valid terms $T$ (i.e. matrices satisfying $P(T) = T$), and forbidden terms $F$ (i.e. matrices satisfying $P(F) = 0$). 

As noted in Appendix A, the allowed terms $T$ are terms of the following forms: $\sigma_\alpha \otimes \sigma_i \otimes \sigma_j \otimes \sigma_0$ (channel, possibly with memory from $A$ to $B$); 
$\sigma_i \otimes \sigma_\beta \otimes \sigma_0 \otimes \sigma_j$ (channel, possibly with memory from $B$ to $A$); $\sigma_\alpha \otimes \sigma_\beta \otimes \sigma_0 \otimes \sigma_0$ (state shared between $A$ and $B$). The projector $P$ is precisely the projector onto the subspace of $\mathcal{L}(\H)$ spanned by the valid terms. Its generalisation for an arbitrary number of parties follows the same logic and can be found in \cite{witness, oreshkov}.

Let now $T = \sigma_\lambda \otimes \sigma_m \otimes \sigma_\nu \otimes \mathbb{1}$. This is a valid term for all $\lambda, \ \nu \, = 0,...,3$ and $m = 1,...,3$. A straightforward calculation gives
\begin{align}
\left[H,T\right] = h_{\alpha \beta \gamma \delta} (& \sigma_\alpha \sigma_\lambda \otimes \sigma_\beta \sigma_m \otimes \left[\sigma_\gamma,\sigma_\nu\right]\otimes \sigma_\delta \nonumber \\ 
+& \sigma_\alpha \sigma_\lambda \otimes \left[\sigma_\beta,\sigma_m \right] \otimes \sigma_\nu \sigma_\gamma \otimes \sigma_\delta 
 \nonumber \\ 
+&  \left[ \sigma_\alpha, \sigma_\lambda \right] \otimes \sigma_m \sigma_\beta \otimes \sigma_\nu \sigma_\gamma \otimes \sigma_\delta 
). 
\end{align}

In order to prove our result, it is useful to extend the definition of $\mathrm{d}_{ijk}$ and $\mathrm{f}_{ijk}$ to greek indices. We set the symbol $\mathrm{f}_{\mu \nu \rho}$ to be equal to the usual structure constant, when the indices run from 1 to $d^2-1$, and to be equal to zero when any index is zero. A similar definition applies for $\mathrm{d}_{ijk}$.  It is then easy to check, with the use of Eq. (\ref{basic}), that
\begin{equation}
\label{trick}
\mathrm{Tr}\left(\left[\sigma_\mu, \sigma_\nu\right] \sigma_\rho\right) = 4\mathrm{i} \, \mathrm{f}_{\mu \nu \rho}. 
\end{equation}
Take now the forbidden term $F = \mathbb{1}\otimes\mathbb{1}\otimes \sigma_n \otimes \sigma_\eta$. Equation (\ref{eqcondition}) then reads
\begin{equation}
\label{constraint}
0 = \mathrm{Tr}\left(\left[H,T\right]F\right) = 32 \mathrm{i} \, h_{\lambda m c \eta} \mathrm{f}_{c \nu n}.
\end{equation}
Because $H$ generates a valid process matrix transformation, Equation (\ref{constraint}) has to be satisfied for all values of $\nu$ and $n$. Note that we have substituted the greek index $\gamma$ with the latin index $c$ because $\mathrm{f}_{\gamma \nu n}$ is trivially zero when $\gamma =0$. Eq. (\ref{constraint}) yields 
\begin{equation}
\label{type1a}
h_{\lambda m n \mu} = 0
\end{equation}
for $\lambda, \ \mu = 0,...,d^2-1$ and $m, \ n = 1,...,d^2-1 $. In order to see that this is the case, we note that the structure constant $f_{ijk}$ is the $j,k$ matrix element of the $i$-th $\mathrm{su}$$(d)$ generator in the adjoint representation, whereby $\mathrm{su}(d) \ni X \, \mapsto \,\mathrm{Ad}_X: \mathrm{su}(d) \longrightarrow \mathrm{su}(d)$, defined by $\mathrm{Ad}_X(Y) = \left[X,Y\right]$, (see, for example, \cite{georgi}). With this fact in mind, we note that Eq. (\ref{constraint}) is just a linear combination of basis elements of $\mathrm{su}$$(d)$. Because the adjoint representation of $\mathrm{su}$$(d)$ is faithful, it preserves the linear independence of the generators. It then follows that the coefficients $h_{\lambda m n \mu}$ must vanish.

Because the linear condition $P(W) = W$ is symmetric in $A$ and $B$, a completely analogous argument to the one leading to Eq. (\ref{type1a}) leads to the same result but with $A$ and $B$ interchanged:
\begin{equation}
\label{type1b}
h_{m \lambda \mu n} = 0.
\end{equation}

Consider now the forbidden term $F = \mathbb{1} \otimes \sigma_\eta \otimes \mathbb{1} \otimes \sigma_n$. Equation (\ref{eqcondition}) reads
\begin{equation}
 32 \mathrm{i} \, h_{\lambda b \nu n} \mathrm{f}_{b m \eta} = 0.	
\end{equation} 
As in the previous case, this leads to 
\begin{equation}
\label{type2a}
h_{\lambda m \nu n} = 0.	
\end{equation}     
By symmetry we also have
\begin{equation}
\label{type2b}
h_{m \lambda n \nu} = 0.	
\end{equation}
Consider now the forbidden term $F = \sigma_\eta\otimes\mathbb{1}\otimes\sigma_n\otimes \sigma_\xi$. Equation (\ref{eqcondition}) reads
\begin{equation}
16 \mathrm{i} \, h_{\alpha m \gamma \xi} (\mathrm{Tr}(\sigma_\alpha \sigma_\lambda \sigma_\eta)\mathrm{f}_{\gamma \nu n} + \mathrm{Tr}(\sigma_\nu \sigma_\gamma \sigma_n) \mathrm{f}_{\alpha \lambda \eta}) = 0.
\end{equation} 
Choosing $\eta = 0$ and $\lambda = l \ \in \{1,...,d^2-1\}$ gives $h_{lmc\xi} = 0$. On the other hand, choosing $\nu = n$ and summing over $n$ yields $h_{lm0\xi} = 0$ after using Eq. (\ref{basic}) and the fact that $\mathrm{d}_{ijk}$ is traceless. We conclude
\begin{equation}
\label{type3}
h_{m n \lambda \mu} = 0.	
\end{equation}
Finally, evaluating Equation (\ref{eqcondition}) for the forbidden term $F = \mathbb{1} \otimes \sigma_\eta \otimes \sigma_\xi \otimes \sigma_n$ leads to $h_{\lambda 0 mn} =0$ for the choice $\eta = m$ (we again use  Eq. (\ref{basic}) and the fact that $\mathrm{d}_{ijk}$ is traceless) and $h_{\lambda i mn} =0$ for $\eta = 0$. We conclude
\begin{equation}
\label{type4}
h_{\lambda \mu  m n} =0	
\end{equation} 
Equations (\ref{type1a}, \ref{type1b}, \ref{type2a}, \ref{type2b}, \ref{type3}, \ref{type4}) imply that $H$ consists only of single-body terms, as we wanted to show. 

At this point it is clear that the argument can be extended to the case of arbitrary parties. In fact, there are only four essentially different types of contributions to $H$ that need to be ruled out: 
\begin{itemize}
\item
Terms in which $h$ has one latin index in some input variable, one latin index in an output variable corresponding to a different party, and greek indices everywhere else.
\item
Terms in which $h$ has one latin index in some input variable, one latin index in an output variable corresponding to the same party, and greek indices everywhere else.
\item
Terms in which $h$ has two latin indices in two input variables and greek indices everywhere else.
\item
Terms in which $h$ has two latin indices in two output variables and greek indices everywhere else.	
\end{itemize}
Each of this type of terms can be ruled out by a natural generalisation of the arguments presented here for the bipartite case. 

\section{$W_{OCB}$ is an extremal process}
\label{appendix2}

In this section we show that the process $W_{OCB}$, given by Equation (\ref{wocb}) in the main text, is extremal. By definition, a process $W \in \mathcal{L}(\H)$, for a given Hilbert space $\H$, is extremal if it cannot be written as a non trivial convex combination of the form $W = qW_1 + (1-q)W_2$, where $W_1$ and $W_2$ are two \textit{different} processes and $0< q < 1$. 

Let $\mathcal{S} = \mathrm{Supp}(W) \subseteq \H$ be the support of $W$ and let $\Pi_\mathcal{S}: \H \longrightarrow \H$ be its corresponding projector. Define the projector $P_\T: \mathcal{L}(\H)\longrightarrow \mathcal{L}(\H)$ by $P_\T(X) = \Pi_\mathcal{S} X \Pi_\mathcal{S}$ for all $X \in \mathcal{L}(\H)$. Here, $\T = \S\otimes \S^\dagger$ denotes the subspace corresponding to the projector $P_\T$. Consider now the projector $P_\V: \mathcal{L}(\H)\longrightarrow \mathcal{L}(\H)$ into the subspace of valid process matrices $V$. This projector is denoted by $P$ in the main text. The fact that $W_{OCB}$ is extremal can be seen as a consequence of the following fact: A process $W$ is extremal if the rank of the projector into the intersection of $T$ and $V$ is equal to 1. Let us prove this fact. Assume that $W$ is not extremal, that is, $W = qW_1 + (1-q)W_2$ for different processes $W_1$ and $W_2$. Let us first show that $\mathrm{Supp}(W_i)\subseteq \S$ for $i = 1,2$. This is equivalent to showing that $\mathrm{Ker}(W)\subseteq\mathrm{Ker}(W_i)$ for $i = 1,2$, where $\mathrm{Ker}(X)$ denotes the Kernel of $X$. Let $\ket{\psi}\in \mathrm{Ker}(W)$ and write $W_i$ in diagonal form, $W_i = \sum_j \lambda^j_i \ketbra{\lambda^j_i}{\lambda^j_i}$, for $i = 1,2$. Since $W\ket{\psi} = 0$, the positivity of the eigenvalues $\lambda ^j_i$ implies $W_i\ket{\psi} = 0$ for $i = 1,2$, meaning that  $\mathrm{Ker}(W)\subseteq\mathrm{Ker}(W_i)$, or, equivalently, $\mathrm{Supp}(W_i)\subseteq \S$, for $i = 1,2$. Therefore, $P_\T(W_i) = W_i$, for $i=1,2$. Moreover, since, by assumption, $W_i$ is a valid process, $P_\V(W_i) = W_i$ for $i = 1,2$. Thus, $W_i \in \T \cap \V$, $i = 1,2$. Because $W_1$ and $W_2$ are different processes, they are linearly independent. This means that $\S \cap \V$ is at least two dimensional. This proves our claim. 

It is then straightforward (with the help of a computer) to build the corresponding projectors for the case of $W_{OCB}$ and to verify that, in this case, $\T \cap \V$ is indeed one dimensional. 
\end{document}